\documentclass[12pt]{article}
\usepackage{amsbsy}
\usepackage{amsfonts}

\begin{document}

\renewcommand\vec[1]{\boldsymbol{#1}}
\def\abel{\vec{\mathcal{A}}}
\def\vzeta{\vec{\zeta}}
\def\veta{\vec{\eta}}

\title{Finite-genus solutions for the Hirota's bilinear difference equation.}

\author{V.E.Vekslerchik
\thanks{Regular Associate of the
   Abdus Salam International Centre for Theoretical Physics,
   Trieste, Italy}
\\
\\
\normalsize\it
  Institute for Radiophysics and Electronics,  \\
\normalsize\it
  National Academy of Sciences of Ukraine,     \\
\normalsize\it
  Proscura Street 12, Kharkov 310085, Ukraine.}

\date{\today}
\maketitle

\begin{abstract}
The finite-genus solutions for the Hirota's bilinear difference equation are
constructed using the Fay's identities for the $\theta$-functions of 
compact Riemann surfaces.
\end{abstract}

In the present work I want to consider once more the question of
constructing the finite-genus solutions for the famous Hirota's
bilinear difference equation (HBDE) \cite{Hirota}

\begin{equation}
\tau(k-1,l,m) \, \tau(k+1,l,m) +
\tau(k,l-1,m) \, \tau(k,l+1,m) +
\tau(k,l,m-1) \, \tau(k,l,m+1)
= 0
\label{hbde}
\end{equation}
which has been solved in \cite{KWZ} using
the so-called algebraic-geometrical approach. This method, which is the
most powerful method for deriving the quasi-periodic solutions (QPS) and
which has been developed for almost all known integrable systems, exploits
some rather sophisticated pieces of the theory of functions of complex
variables and is based on some theorems determining the number of
functions with prescribed structure of singularities on the Riemann
surfaces (see \cite{Krichever} for review).  However, in some cases the
QPS can be found with less efforts, in a more straightforward way, using
the fact that the finite-genus QPS (and namely they are the subject of this
note) of all integrable equations possess similar and rather simple
structure: up to some phases they are 'meromorphic' combinations of the
$\theta$-functions associated with compact Riemann surfaces \cite{Mumford}
(the situation resembles the pure soliton case, where solutions are
rational functions of exponents). Thus, to construct these solutions we
only have to determine some constant parameters.  This, as in the pure
soliton case, can be done directly, using the well-known properties of the
$\theta$-functions of compact Riemann surfaces. In \cite{V99} such
approach was developed for the Ablowitz-Ladik hierarchy, where the
finite-genus solutions were 'extracted' from the so-called Fay's formulae
\cite{Fay,Mumford}. The fact, that the HBDE is closely related to the
Fay's identities is not new and was mentioned, e.g., in \cite{KWZ} (see
Remark 2.7), but contrary to this work I will use these identities as a
starting point and will show how one can derive from them, by rather short
and very simple calculations (which to my knowledge have not been
presented explicitly in the literature), a wide family of solutions for
the HBDE.

Consider a compact Riemann surface $X$ of the genus $g$.
One can choose a set of closed contours (cycles) $\{ a_{i}, b_{i} \}_{i=1,
..., g}$ with the intersection indices

\begin{equation}
a_{i} \circ a_{j} =
b_{i} \circ b_{j} = 0,
\qquad
a_{i} \circ b_{j} = \delta_{ij}
\qquad
i,j = 1, \dots, g
\end{equation}
and find $g$ independent holomorphic differentials which
satisfy the normalization conditions

\begin{equation}
\oint_{a_{i}} \omega_{k} = \delta_{ik}
\end{equation}
The matrix of the $b$-periods,

\begin{equation}
\Omega_{ik} = \oint_{b_{i}} \omega_{k}
\end{equation}
determines the so-called period lattice,
$L_{\Omega} = \left\{
   \vec{m} + \Omega \vec{n},
   \quad
   \vec{m}, \vec{n} \in {\mathbb Z}^{g}
   \right\}$,
the Jacobian of this surface
$\mathrm{Jac}(X)={\mathbb C}^{g}/L_{\Omega}$ (2$g$ torus) and the Abel
mapping $X \to \mathrm{Jac}(X)$,

\begin{equation}
\abel(P) = \int^{P}_{P_{0}} \vec{\omega}
\label{abel}
\end{equation}
where $\vec{\omega}$ is the $g$-vector of the 1-forms,
$\vec{\omega} =
\left( \omega_{1}, \dots, \omega_{g} \right)^{\scriptscriptstyle T}$
and $P_{0}$ is some fixed point of $X$.

A central object of the theory of the compact Riemann surfaces is the
$\theta$-function, $\theta(\vzeta)=\theta(\vzeta,\Omega)$,

\begin{equation}
\theta\left(\vzeta\right) =
\sum_{ \vec{n} \, \in \, {\mathbb Z}^{g} }
\exp\left\{
     \pi i \, \left( \vec{n}, \Omega \vec{n} \right) \; +
   2 \pi i \, \left( \vec{n}, \vzeta \right)
\right\}
\end{equation}
where $( \vec{\xi},\vec{\eta} )$ stands for $\sum_{i=1}^{g}
\xi_{i}\eta_{i}$, which is a quasiperiodic function on $\mathbb{C}^{g}$

\begin{eqnarray}
\theta\left(\vzeta + \vec{n}\right) &=&
   \theta\left(\vzeta\right)
\\
\theta\left(\vzeta + \Omega\vec{n}\right) &=&
   \exp\left\{
      -   \pi i \, \left( \vec{n}, \Omega \vec{n} \right) \;
      - 2 \pi i \, \left( \vec{n}, \vzeta \right)
   \right\}
\theta\left(\vzeta\right)
\end{eqnarray}
for any $\vec{n} \in {\mathbb Z}^{g}$.

The famous Fay's trisecant formula can be written as

\begin{equation}
\sum_{i=1}^{3}
a_{i} \;
\theta\left( \vzeta + \veta_{i} \right)
\theta\left( \vzeta - \veta_{i} \right)
= 0
\label{fay}
\end{equation}
where

\begin{eqnarray}
2 \veta_{1} &=&
  - \abel(P_{1}) + \abel(P_{2}) + \abel(P_{3}) - \abel(P_{4}),
\\
2 \veta_{2} &=&
    \phantom{-}
    \abel(P_{1}) - \abel(P_{2}) + \abel(P_{3}) - \abel(P_{4}),
\\
2 \veta_{3} &=&
    \phantom{-}
    \abel(P_{1}) + \abel(P_{2}) - \abel(P_{3}) - \abel(P_{4}).
\end{eqnarray}
Here $P_{1}, ..., P_{4}$ are arbitrary points of $X$, and the constants
$a_{i}$ are given by

\begin{eqnarray}
a_{1} &=& e(P_{4},P_{1}) \; e(P_{2},P_{3}), \\
a_{2} &=& e(P_{4},P_{2}) \; e(P_{3},P_{1}), \\
a_{1} &=& e(P_{4},P_{3}) \; e(P_{1},P_{2}).
\end{eqnarray}
The skew-symmetric function $e(P,Q)$, $e(P,Q)=-e(Q,P)$, is closely related
to the prime form \cite{Mumford} and is given by

\begin{equation}
e(P,Q) =
\theta
\left[ \vec{\delta}', \vec{\delta}'' \right]
\left( \abel(Q) - \abel(P) \right)
\label{pf}
\end{equation}
where
$\theta \left[ \vec{\alpha}, \vec{\beta} \right] \left( \vzeta \right)$
is the so-called $\theta$-function with characteristics,

\begin{equation}
\theta \left[ \vec{\alpha}, \vec{\beta} \right]
\left( \vzeta \right) =
\exp\left\{
     \pi i \, \left( \vec{\alpha}, \Omega \vec{\alpha} \right) +
   2 \pi i \, \left( \vec{\alpha}, \vzeta + \vec{\beta}\right)
\right\}
\theta \left( \vzeta + \Omega \vec{\alpha} + \vec{\beta} \right),
\end{equation}
and
$\left( \vec{\delta}', \vec{\delta}'' \right)
\in {1 \over 2} \mathbb{Z}^{2g} / \mathbb{Z}^{2g}$ is a non-singular odd
characteristics,

\begin{equation}
\theta \left[ \vec{\delta}', \vec{\delta}'' \right]
\left( \vec{0} \right) = 0,
\qquad
\mathrm{grad}_{\vzeta} \;
\theta \left[ \vec{\delta}', \vec{\delta}'' \right]
\left( \vec{0} \right) \ne \vec{0}
\end{equation}

Now it is very easy to establish relations between (\ref{fay}) and
the HBDE (\ref{hbde}). To do this one has first to introduce the
discrete variables by

\begin{equation}
\Theta(k,l,m) =
\theta\left( \vzeta + k \veta_{1} + l \veta_{2} + m \veta_{3} \right)
\label{step_one}
\end{equation}
The Fay's identity (\ref{fay}) can now be rewritten as

\begin{eqnarray}
a_{1} \, \Theta(k-1,l,m) \, \Theta(k+1,l,m)
+
a_{2} \, \Theta(k,l-1,m) \, \Theta(k,l+1,m)
\cr
+
a_{3} \, \Theta(k,l,m-1) \, \Theta(k,l,m+1)
&=& 0
\end{eqnarray}
from which it follows that the quantity

\begin{equation}
\tau(k,l,m) =
a_{1}^{k^{2}/2} \,
a_{2}^{l^{2}/2} \,
a_{3}^{m^{2}/2} \;
\Theta(k,l,m)
\label{step_two}
\end{equation}
satisfies HBDE (\ref{hbde}).

Thus the last formula,

\begin{equation}
\tau(k,l,m) =
a_{1}^{k^{2}/2} \,
a_{2}^{l^{2}/2} \,
a_{3}^{m^{2}/2} \;
\theta\left( \vzeta + k \veta_{1} + l \veta_{2} + m \veta_{3} \right)
\label{solution}
\end{equation}
where $\theta$ is the $\theta$-function of some compact Riemann surface
$X$ and the constants $a_{i}$ and $\eta_{i}$ depend on four points
(paths) on this surface, determines a family of finite-genus solutions of
the HBDE.

Now I want to discuss the solutions obtained above. First of all it should
be noted that these solutions are 'finite-genus' but not quasiperiodic. At
first glance, this seems to be strange, because usually the finite-genus
solutions naturally appear when one solves quasiperiodic problems. For,
example, in $1+1$ dimensional discrete systems, such as Toda chain,
Ablowitz-Ladik equations, etc, the quasiperiodicity leads to the
polynomial dependence of the scattering matrix of the auxiliary problem on
the spectral parameter. These polynomials determine some hyperelliptic
curve (spectral curve) of finite genus, and the quasiperiodic solutions
are built up of the $\theta$-functions corresponding to the latter. Thus,
in some sense, in discrete systems the quasiperiodicity implies the
'finite-genus' property.  In other words, quasiperiodic solutions are
finite-genus.  However, the reverse is not obligatory true. The Fay's
identities do not imply that the integrals $\abel(P)$ are in some way
related to the periods $\oint\vec{\omega}$. Formula (\ref{fay}) determine,
so to say, 'local' properties of the $\theta$-functions, and all above
consideration was local, without reference to some boundary conditions
(quasiperiodicity).

Another point which I would like to discuss here is to compare the approach
of this note with the algebro-geometrical one. The ideology of the latter
is to operate on the Riemann surface: the key moment in calculating the
Baker-Akhiezer function, the central object of the algebro-geometrical
method, is to satisfy the condition that it is a single-valued function of
the point of the Riemann surface. Since we didn't introduce the
Baker-Akhiezer function and didn't study its analytical properties the
question of how our solutions depend on the points $P_{i} \in X$ (or on
the integral paths from $P_{0}$ to $P_{i}$, to be more precise) is not so
crucial as in the algebro-geometrical approach and is in some sense the
question of parametrization of constants. One can restrict the points
$P_{i}$, $i=1,2,3$, to some neighborhood of the point $P_{0}$ and rewrite
the Abel's integrals in terms of local coordinates (with some polynomial
representing the Riemann surface). This enables not to mention the Riemann
surface and reformulate all results in terms of integrals over the complex
plane. As to the 'global' (or 'homotopical') effects, which arise when we
add to the paths $(P_{0},P_{i})$ some integer cycles
($\sum_{k=1}^{g} m_{k} a_{k} + n_{k} b_{k}$,
$m_{k}, n_{k} \in {\mathbb Z}$) it should be noted that,
if we consider the Abel integral as mapping
$X \to {\mathbb C}^{g}$, then such deformations of the contours
change $\abel(P)$ as
$\abel(P) \to \abel(P) + \vec{\gamma}$,
$\vec{\gamma} \in L_{\Omega}$.
This results first in adding some {\it half-period} to the argument of the
$\theta$-function in (\ref{solution}),

\begin{equation}
\vzeta + \sum_{i=1}^{3} k_{i} \veta_{i}
\to
\vzeta + \sum_{i=1}^{3} k_{i} \veta_{i} +
{1 \over 2} \sum_{i=1}^{3} k_{i} \vec{\gamma_{i}},
\qquad
\vec{\gamma_{i}} \in L_{\Omega}
\end{equation}
and, second, in altering the constants $a_{i}$'s,
$a_{i} \to a_{i} \exp(f_{i})$.
Thus, we come to the point where one can try to apply the theory of
Backlund transformations for the HBDE to describe the transformations of
$\theta$-functions due to half-period shifts (and vice versa). This is an
interesting problem, which deserves special studies.

To conclude I would like to note the following. In principle, the direct
approach based on the Fay's identities can be used to derive the
finite-genus solutions not only for the HBDE but for almost all known
integrable systems (some examples one can find in the book
\cite{Mumford}). However, contrary to the case of the HBDE where all
'calculations' take only two lines (formulae (\ref{step_one}) and
(\ref{step_two}) above), in the case of partial differential equations
such as, e.g., KP equation the corresponding calculations become rather
cumbersome. For example, to solve the KP equation one has to expand the
Fay's identities up to the third order in some small parameter. From the
other hand, it is a widely known fact that almost all known integrable
equations can be derived from the HBDE. Hence one can 'skip' the Fay's
identities and use solutions (\ref{solution}) as a starting point. In
\cite{V99} it was shown how to obtain some finite-genus solutions for the
nonlinear Schr\"{o}dinger and KP equations using the corresponding
solutions for the Ablowitz-Ladik hierarchy, which can be viewed as some
'pre-continuous' version of the HBDE (in the Ablowitz-Ladik hierarchy 2 of
3 discrete coordinates of the HBDE are presented as the Miwa's shifts of
two infinite sets of continuous variables).

%%%%%%%%%%%%%%%%%%%%%%%%%%%%%

\end{document}